\documentclass[pra,aps,showpacs,superscriptaddress,twocolumn]{revtex4}

\usepackage{amsmath}
\usepackage{bm}
\usepackage{graphicx}

\begin{document}

\title{Symmetry breaking Rayleigh-Taylor instability in a two-component
Bose-Einstein condensate}

\author{Tsuyoshi Kadokura}
\affiliation{Department of Engineering Science, University of
Electro-Communications, Tokyo 182-8585, Japan}

\author{Tomohiko Aioi}
\affiliation{Department of Engineering Science, University of
Electro-Communications, Tokyo 182-8585, Japan}

\author{Kazuki Sasaki}
\affiliation{Department of Engineering Science, University of
Electro-Communications, Tokyo 182-8585, Japan}

\author{Tetsuo Kishimoto}
\affiliation{Department of Engineering Science, University of
Electro-Communications, Tokyo 182-8585, Japan}
\affiliation{Center for Frontier Science and Engineering, University of
Electro-Communications, Tokyo 182-8585, Japan}
\affiliation{PRESTO, Japan Science an Technology Agency (JST), Saitama
	332-0012, Japan}

\author{Hiroki Saito}
\affiliation{Department of Engineering Science, University of
Electro-Communications, Tokyo 182-8585, Japan}

\date{\today}

\begin{abstract}
The interfacial instability and subsequent dynamics in a phase-separated
two-component Bose-Einstein condensate with rotation symmetry are
studied.
When the interatomic interaction or the trap frequency is changed, the
Rayleigh-Taylor instability breaks the rotation symmetry of the interface,
which is subsequently deformed into nonlinear patterns including mushroom
shapes.
\end{abstract}


\maketitle

\section{Introduction}

The Rayleigh-Taylor instability~\cite{Rayleigh,Taylor,Lewis,Chandra} (RTI)
is an instability of an interface between two fluids in a metastable
state.
For instance, when a layer of a heavier fluid is laid on a lighter fluid,
the system is energetically unfavorable and the two fluids tend to exchange
their positions.
However, if the two fluids are immiscible and their interface is flat, the
exchange cannot occur without breaking the translation symmetry of the
interface.
Once an infinitesimal modulation arises on the interface, it exponentially
grows due to the RTI, and the interface develops into complicated patterns
such as a mushroom-shaped pattern.
This kind of phenomena is found through nature in a wide scale ranging
from laboratory to astronomical scales~\cite{Schmidt}.

Recently, the RTI was predicted to be observed in a two-component
Bose-Einstein condensate (BEC)~\cite{SasakiRTI,Gautam}.
In Ref.~\cite{SasakiRTI}, a two-component BEC trapped in a tight
pancake-shaped trap is considered, in which the two components separate
into two semicircular shapes in the trap.
When an external force is applied to each component in the direction to
the other component, the RTI sets in and the interface deforms into a
mushroom-shaped pattern.
In Ref.~\cite{Gautam}, the initial state of a two-component BEC forms a
domain structure in the axial direction of a cigar-shaped trap.
The interaction between atoms in one component is then increased, and the
RTI arises at the domain wall.

The RTI is a symmetry breaking phenomenon, that is, even when the
interface has symmetry (e.g. the translation symmetry of a flat interface)
an infinitesimal modulation grows exponentially and the symmetry is
spontaneously broken.
However, in the above mentioned studies in Refs.~\cite{SasakiRTI,Gautam},
the symmetry breaking cannot be observed explicitly, since the relevant
symmetry is broken from the initial state:
the interfaces have edges or rims and have no translation symmetry.
The inhomogeneous interfaces of the initial state strongly affects the
deformation dynamics of the interfaces, obscuring the symmetry breaking
nature of the RTI.

In this paper, we propose a system to observe the symmetry breaking
RTI for a trapped two-component BEC.
We consider a two-component BEC with rotation symmetry, in which the two
components separate radially and a ``bubble'' of inner component is
surrounded by a shell of outer component.
The interface between the two components has a spherical shape for an
isotropic trap and a circular shape for a quasi-2D axisymmetric trap.
If we change a parameter in such a way that the inner component tends to
go out of the outer shell component, the RTI breaks the rotation symmetry
of the interface and the spherical or circular interface is deformed into
various patterns.
The symmetry breaking RTI can thus be realized in a trapped BEC.
The RTI that breaks rotation symmetry occurs in a variety of systems, such
as supernova explosion~\cite{Burrows}, imploding targets in
inertial-confinement fusion~\cite{Sakagami}, and collapsing cavitation
bubbles~\cite{Plesset,Brenner}.

This paper is organized as follows.
Section~\ref{s:formulation} provides a formulation of the problem and 
Sec.~\ref{s:numerical} shows numerical results.
Section~\ref{s:2d} demonstrates the symmetry-breaking RTI and subsequent
dynamics for an axisymmetric oblate system.
Section~\ref{s:3d} shows dynamics for an isotropic trap and performs
Bogoliubov analysis.
Section~\ref{s:osci} examines oscillation of interaction.
Section~\ref{s:conc} gives conclusions to this study.

\section{Formulation of the problem}
\label{s:formulation}

We consider a mixture of two kinds of bosonic atoms with mass $m_1$ and
$m_2$ confined in trapping potential $V_1$ and $V_2$, respectively.
The Hamiltonian for the system is given by
\begin{eqnarray}
\hat H & = & \sum_{j=1}^2 \int d\bm{r} \hat\psi_j^\dagger \left(
-\frac{\hbar^2}{2m_j} \nabla^2 + V_j \right) \hat\psi_j
\nonumber \\
& & + \sum_{j,j'} \int
d\bm{r} d\bm{r}' \hat\psi_j^\dagger (\bm{r}) \hat\psi_{j'}^\dagger(\bm{r}')
U_{jj'}(\bm{r} - \bm{r}') \hat\psi_{j'}(\bm{r}') \hat\psi_j(\bm{r}),
\nonumber \\
\end{eqnarray}
where $\hat\psi_j$ is the bosonic field operator for component $j$ and
$U_{jj'}$ is the interaction between atoms of components $j$ and $j'$.
In the mean-field theory, we assume that the atoms in each component
occupy the same wave function $\psi_j$ and that the interaction potential
is reduced to the Fermi pseudopotential,
\begin{eqnarray}
U_{jj'}(\bm{r} - \bm{r}') & = & 2 \pi \hbar^2 a_{jj'} \left( m_j^{-1}
+ m_{j'}^{-1} \right) \delta(\bm{r} - \bm{r}') \nonumber \\
& \equiv & g_{jj'} \delta(\bm{r} - \bm{r}'),
\end{eqnarray}
where $a_{jj'}$ is the $s$-wave scattering length between the atoms in
components $j$ and $j'$.
The system is thus described by the two-component Gross-Pitaevskii (GP)
equation $(j \neq j')$,
\begin{equation} \label{GP}
i\hbar \frac{\partial \psi_j}{\partial t} = \left( -\frac{\hbar^2}{2m_j}
\nabla^2 + V_j + g_{jj} |\psi_j|^2 + g_{jj'} |\psi_{j'}|^2 \right)
\psi_j.
\end{equation}
The macroscopic wave functions are normalized as $\int |\psi_j|^2 d\bm{r}
= N_j$ with $N_j$ being the number of atoms in component $j$.
The two components are miscible for $g_{11} g_{22} > g_{12}^2$ and
immiscible for $g_{11} g_{22} < g_{12}^2$.

We solve the 3D GP equation (\ref{GP}) numerically using the
pseudospectral method~\cite{Recipes}.
The initial state is the ground state prepared by the imaginary-time
propagation method, in which $i$ on the left-hand side of Eq.~(\ref{GP})
is replaced by $-1$.
We then add a small noise to the initial state as a seed that triggers the
RTI.
The dynamics do not depend on the initial noise qualitatively.

In the following calculations, we assume a dual-species BEC with
$^{85}{\rm Rb}$ and $^{87}{\rm Rb}$, where the $|f = 2, m_f = -2 \rangle$
state of $^{85}{\rm Rb}$ is component 1 and the $|f = 1, m_f = -1 \rangle$
state of $^{87}{\rm Rb}$ is component 2.
This system has been realized by the JILA group~\cite{Papp}, in which
controlled phase separation was observed by changing the $s$-wave
scattering length $a_{11}$ of $^{85}{\rm Rb}$ using a magnetic-field
Feshbach resonance, which is variable in the range $a_{11} = 50$-$900
a_{\rm B}$ with $a_{\rm B}$ being the Bohr radius.
Since $a_{22} = 99 a_{\rm B}$ and $a_{12} = 213 a_{\rm B}$, the condition
for the phase separation is satisfied for $a_{11} < 458 a_{\rm B}$.

\section{Numerical results}
\label{s:numerical}

\subsection{Rayleigh-Taylor instability in axisymmetric oblate systems}
\label{s:2d}

\begin{figure}[t]
\includegraphics[width=8cm]{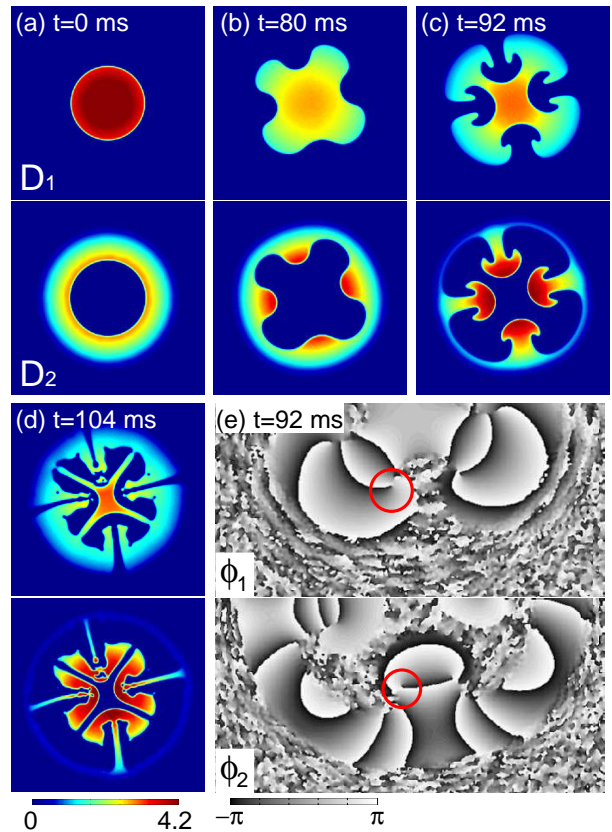}
\caption{
(Color online) (a)-(d) Dynamics of the column density profiles $D_1 = \int
|\psi_1|^2 dz$ (upper panels) and $D_2 = \int |\psi_2|^2 dz$ (lower
panels) in an axisymmetric trap with $(\omega_\perp, \omega_z) = 2\pi
\times (25, 1250)$ Hz.
The scattering length $a_{11}$ is linearly increased from $80 a_{\rm B}$
to $240 a_{\rm B}$ between $t = 0$ and $t = 40$ ms, and after that
$a_{11}$ is fixed to $240 a_{\rm B}$.
The numbers of atoms are $N_1 = N_2 = 10^5$.
The unit of the column density is $10^{12}$ ${\rm cm}^{-2}$.
(e) Cross-sectional phase profile $\phi_j = {\rm arg} [\psi_j(z = 0)]$ of
the lower half region of (c).
The circles in (e) indicate examples of quantized vortices created under
the caps of the mushrooms.
The field of view is $65.4 \times 65.4$ $\mu{\rm m}$ in (a)-(d) and $65.4
\times 32.7$ $\mu{\rm m}$ in (e).
}
\label{f:pancake}
\end{figure}
We first demonstrate the dynamics for an axisymmetric oblate trap,
$V_j = m_j [\omega_\perp^2 (x^2 + y^2) + \omega_z^2 z^2] / 2$,
where $\omega_z \gg \omega_\perp$.
We assume that the gravitational sag is compensated and the two components
share a common trap center.
Figure~\ref{f:pancake} shows the time evolution of the density and phase
profiles of the system, obtained by solving the 3D GP equation (\ref{GP}).
The initial state is the ground state for $a_{11} = 80 a_{\rm B}$ and $N_1
= N_2$, which has the axisymmetric circular interface between the two
components [Fig.~\ref{f:pancake} (a)].
The repulsive interaction of component 1 (inner) is then gradually
increased and when it exceeds that of component 2 (outer) the system
becomes metastable, i.e., the state in which component 1 surrounds
component 2 becomes energetically favorable.
At $t \simeq 80$ ms, the axisymmetry of the system is broken and the
interface is modulated due to the RTI [Fig.~\ref{f:pancake} (b)].
The modulation of the interface subsequently grows to become a four-fold
mushroom shape [Fig.~\ref{f:pancake} (c)].
Quantized vortices are generated under the caps of the mushrooms in both
components [circles in Fig.~\ref{f:pancake} (e)].
When the tops of the mushrooms reach the edge or the center of the system,
a highly nonlinear pattern is observed [Fig.~\ref{f:pancake} (d)].
The $n$-fold mushroom shapes with $n \neq 4$ are also observed, where $n$
is larger for a larger final value of $a_{11}$.

\begin{figure}[t]
\includegraphics[width=8.5cm]{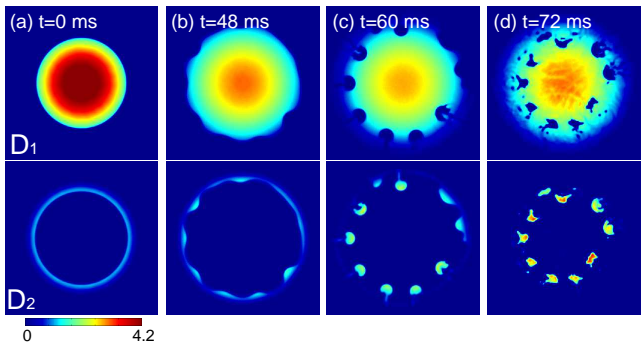}
\caption{
(Color online) Dynamics of the column density profiles $D_1$ and $D_2$ for
$N_1 = 1.8 \times 10^5$ and $N_2 = 2 \times 10^4$.
Other parameters are the same as those in Fig.~\ref{f:pancake}.
}
\label{f:droplet}
\end{figure}
The dynamics also depends on the ratio between the numbers of atoms $N_2 /
N_1$.
Figure~\ref{f:droplet} shows the dynamics for $N_2 / N_1 = 1 / 9$.
After the repulsive interaction of component 1 is increased, the RTI
causes modulation at the interface [Fig.~\ref{f:droplet} (b)].
Since $N_2$ is small, the ring of component 2 splits into droplets, which
enter the component 1 forming small mushrooms [Fig.~\ref{f:droplet} (c)].
The droplets of component 2 then go towards the center and gather, where
their complicated shapes are similar to air bubbles rising in water.

\subsection{Rayleigh-Taylor instability in an isotropic system}
\label{s:3d}

\begin{figure}[t]
\includegraphics[width=8.5cm]{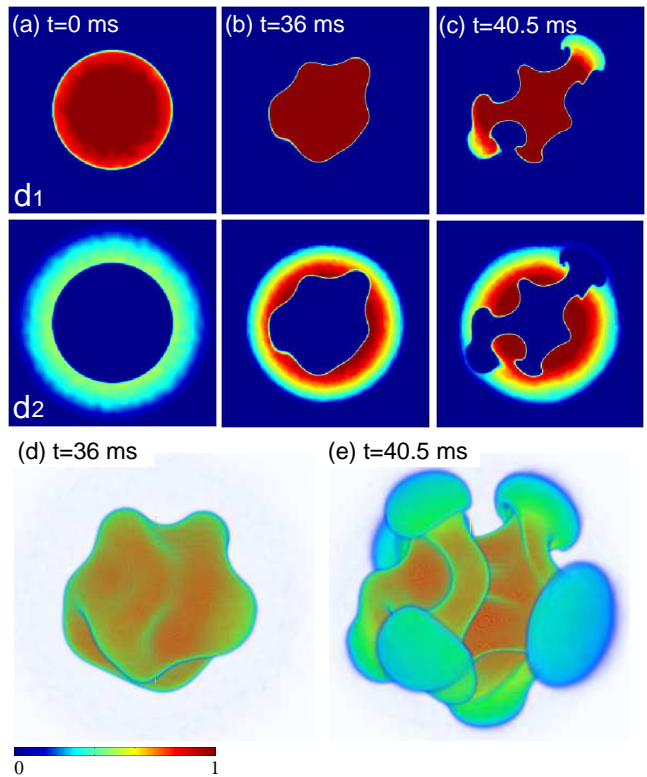}
\caption{
(Color online) (a)-(c) Dynamics of the cross-sectional density profiles
$d_1 = |\psi_1(z = 0)|^2$ and $d_2 = |\psi_2(z = 0)|^2$ of components 1
and 2 and (d), (e) the isodensity surfaces of component 1 in an isotropic
trap with frequency $\omega_1(t = 0) = \omega_2 = 2\pi \times 33.3$ Hz.
The trap frequency $\omega_1$ is increased such that $\omega_1^2$ is
linearly increased from $(\omega_1 / \omega_2)^2 = 1$ to $3$ between $t =
0$ and $t = 30$ ms, and after that $(\omega_1 / \omega_2)^2$ is fixed to
3.
The scattering length of component 1 is $a_{11} = 200 a_{\rm B}$ and the
numbers of atoms are $N_1 = N_2 = 5.2 \times 10^6$.
The unit of the density is $3.0 \times 10^{14}$ ${\rm cm}^{-3}$.
The field of view of each panel is $56.6 \times 56.6$ $\mu{\rm m}$.
}
\label{f:sphere}
\end{figure}
Next we consider a system confined in an isotropic trap given by $V_j =
m_j \omega_j^2 r^2 / 2$ with $r^2 = x^2 + y^2 + z^2$.
The initial state is the ground state for $a_{11} = 200 a_{\rm B}$ and
$\omega_1 = \omega_2$, in which component 2 with a spherical shape is
surrounded by a shell of component 1 [Fig.~\ref{f:sphere} (a)].
The trap frequency $\omega_1$ of component 1 is then increased gradually.
The outer component is pushed inward by the increase in the trap frequency
and the RTI is induced at the spherical interface.
At $t \simeq 36$ ms, the RTI breaks the rotation symmetry and the
spherical interface is modulated [Figs.~\ref{f:sphere} (b) and
\ref{f:sphere} (d)].
The interface is then deformed into a ``mushroom ball''
[Fig.~\ref{f:sphere} (e)].

The unstable modes of the interface is estimated by a simple analysis.
We assume inviscid, incompressible, and irrotational fluids and component
2 of a spherical bubble with radius $R$ is surrounded by component 1.
The excitation frequency $\Omega$ of the interfacial mode proportional to
the spherical harmonics $Y_l^m(\theta, \phi)$ is given by~\cite{Lamb}
\begin{eqnarray} \label{analytic}
\Omega^2 & = & \frac{l (l + 1)}{R [l m_1 n_1 + (l + 1) m_2 n_2]}
\nonumber \\
& & \times \left[ n_2 f_2 - n_1 f_1 + \frac{(l - 1)(l + 2)}{R^2} \sigma \right],
\end{eqnarray}
where $n_j$ is the atomic density, $f_j$ is the external force acting on
an atom at the interface, and $\sigma$ is the interfacial tension
coefficient.
If $\Omega$ is pure imaginary, i.e., the right-hand side of
Eq.~(\ref{analytic}) is negative, the mode is dynamically unstable.
Using the expression of $\sigma$ for a two-component BEC derived in
Ref.~\cite{Schae} and $f_j = m_j \omega_j^2 R$, we find that the modes for
$1 \leq l \leq 7$ are unstable for the parameters in Fig.~\ref{f:sphere},
which seems to be consistent with the interfacial pattern shown in
Fig.~\ref{f:sphere}.
However, this estimation is only qualitative since the compressibility and
the inhomogeneous density distribution of a BEC are not taken into
account.

From Eq.~(\ref{analytic}), we find that the RTI is induced by an increase
in $\rho_1$ or $f_1$, or by a decrease in $\rho_2$ or $f_2$.
The density $\rho_j$ depends on the interaction: an increase (decrease) in
$a_{jj}$ expands (contracts) component $j$, which decreases (increases)
$\rho_j$.
The force $f_j$ acting on each component can be controlled, if the
external trapping potential for each component can be controlled
independently.
The RTI can thus be induced in several ways:
for example, (i) an increase in the scattering length of
inner component, (ii) a decrease in the trap frequency for inner
component, (iii) a decrease in the scattering length of outer component,
and (iv) an increase in the trap frequency for outer component.
The dynamics shown in Figs.~\ref{f:pancake} and \ref{f:sphere} correspond
to (i) and (iv), respectively.
We have numerically confirmed that the RTI can be observed by all the
methods (i)-(iv) for both axisymmetric oblate trap and isotropic trap.

For more precise understanding of the instability, we perform the
Bogoliubov analysis for an isotropic trap.
We expand the GP equation (\ref{GP}) up to the first order of the
deviation $\delta \psi_j(\bm{r})$ from the initial stationary state
$\Psi_j(r)$ with spherical symmetry. 
The excitation mode of the form
\begin{equation} \label{Ylm}
\delta \psi_j = u_j(r) Y_l^m(\theta, \phi) + v_j^*(r)
Y_l^{m*}(\theta, \phi)
\end{equation}
obeys the Bogoliubov-de Gennes equation $(j \neq j')$
\begin{subequations} \label{BdG}
\begin{eqnarray}
& & \left( K_{jl} + V_j - \mu_j + 2 g_{jj} \Psi_j^2 + g_{jj'} \Psi_{j'}^2
\right) u_j + g_{jj} \Psi_j^2 v_j
\nonumber \\
& & + g_{jj'} \Psi_j \Psi_{j'} (u_{j'} +
v_{j'}) = \hbar \Omega u_j,
\\
& & \left( K_{jl} + V_j - \mu_j + 2 g_{jj} \Psi_j^2 + g_{jj'} \Psi_{j'}^2
\right) v_j + g_{jj} \Psi_j^2 u_j 
\nonumber \\
& & + g_{jj'} \Psi_j \Psi_{j'} (u_{j'} +
v_{j'}) = -\hbar \Omega v_j,
\end{eqnarray}
\end{subequations}
where $\mu_j$ is the chemical potential and
\begin{equation}
K_{jl} = -\frac{\hbar^2}{2m_j} \left[ \frac{d^2}{dr^2} + \frac{2}{r}
	\frac{d}{dr} - \frac{l (l + 1)}{r^2} \right].
\end{equation}
The stationary wave function $\Psi_j$ is assumed to be real without loss
of generality.
\begin{figure}[t]
\includegraphics[width=8.5cm]{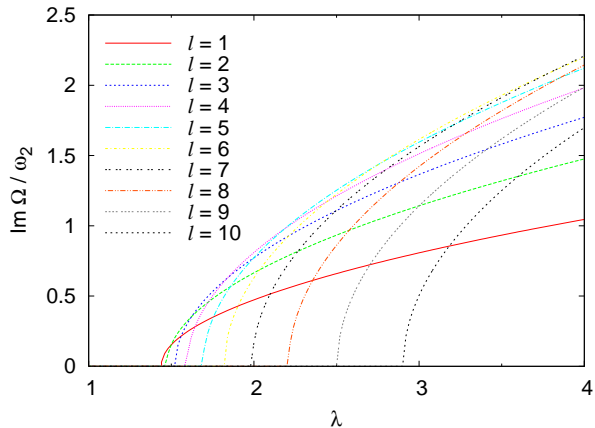}
\caption{
(Color online) Imaginary part of the Bogoliubov excitation frequency, Im
$\Omega$, as a function of $\lambda \equiv (\omega_1 / \omega_2)^2$.
The parameters are the same as those in Fig.~\ref{f:sphere}.
The modes for $l \leq 8$ are plotted, where $l$ is defined in
Eq.~(\ref{Ylm}).
}
\label{f:bogo}
\end{figure}
We numerically diagonalize Eq.~(\ref{BdG}) to study the stability of the
system.
If there is a complex frequency $\Omega$, the corresponding mode grows
exponentially and the system is dynamically unstable.
Figure~\ref{f:bogo} shows the imaginary part of the Bogoliubov excitation
frequency, Im $\Omega$, as a function of $(\omega_1 / \omega_2)^2$.
The critical value of $(\omega_1 / \omega_2)^2$ above which Im $\Omega$
rises increases with an increase in $l$, and above this critical value of
$(\omega_1 / \omega_2)^2$, Im $\Omega$ monotonically increases.
At $(\omega_1 / \omega_2)^2 = 3$, which corresponds to
Fig.~\ref{f:sphere}, Im $\Omega$ for the $l = 5$-$7$ modes are comparably
large and these modes dominate the unstable dynamics.

\subsection{Oscillation of the interaction}
\label{s:osci}

\begin{figure}[t]
\includegraphics[width=8.5cm]{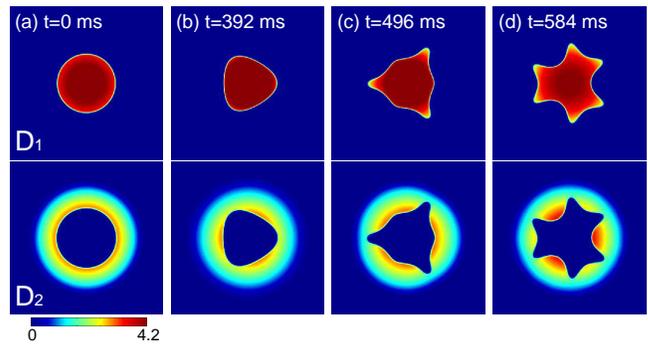}
\caption{
(Color online) Dynamics of the column density profiles $D_1$ and $D_2$,
in which $a_{11}$ is oscillated as in Eq.~(\ref{osc}) with $a_0 = 80
a_{\rm B}$, $A = 0.4$, and $\omega_a = 2 \pi \times 30$ Hz.
Other parameters are the same as those in Fig.~\ref{f:pancake}.
The unit of the column density is $10^{12}$ ${\rm cm}^{-2}$.
The field of view of each panel is $65.4 \times 65.4$ $\mu{\rm m}$.
}
\label{f:oscillate}
\end{figure}
We demonstrate another symmetry breaking dynamics that is not due to the
RTI.
We return to the tight pancake-shaped trap used in Fig.~\ref{f:pancake},
and study the dynamics for an oscillating interaction.
The scattering length of component 1 is oscillated as
\begin{equation} \label{osc}
a_{11} = a_0 (1 + A \sin \omega_a t),
\end{equation}
where $a_0$, $A$, and $\omega_a$ are constants.
Figure~\ref{f:oscillate} shows the dynamics of the column density profile.
By the oscillating repulsive interaction of the inner component, the
circular interface undergoes breathing oscillation.
The axisymmetry of the interface is then spontaneously broken
[Fig.~\ref{f:oscillate} (b)], which is understood as the parametric
amplification of an interface mode.
The interface subsequently exhibits various patterns
[Figs.~\ref{f:oscillate} (b) and \ref{f:oscillate} (c)].
A variety of patterns can be observed depending on the parameters in
Eq.~(\ref{osc}) (data not shown).

\section{Conclusions}
\label{s:conc}

In conclusion, we have investigated the interfacial instabilities and
subsequent dynamics in phase-separated two-component BECs.
Since the initial state has rotation symmetry, the symmetry breaking
nature of the RTI can specifically be observed in this system.
We have demonstrated the RTI and ensuing dynamics for an axisymmetric
oblate trap (Figs.~\ref{f:pancake} and \ref{f:droplet}) and an isotropic
trap (Fig.~\ref{f:sphere}), and the mushroom-shaped patterns are observed
for both systems breaking the rotation symmetry.
We performed the Bogoliubov analysis for the isotropic system and obtained
the unstable spectrum (Fig.~\ref{f:bogo}).
We also examined the dynamics for oscillating interaction and found that
the axisymmetry is spontaneously broken and nonlinear patterns emerge
(Fig.~\ref{f:oscillate}).

In view of the recent development in the control of two-component
BECs~\cite{Mertes,Papp,Tojo}, we expect that not only the phenomena
predicted in the present paper but also other theoretical
predictions~\cite{Saito,Takeuchi,Suzuki,Bezett,Koby,SasakiPRI} concerning
the interfacial instabilities in two-component BECs will be realized in
experiments in the near future.

\begin{acknowledgments}
This work was supported by Grants-in-Aid for Scientific
Research (No.\ 22340116, No.\ 23540464, and No.\ 23740307) from the
Ministry of Education, Culture, Sports, Science and Technology of Japan,
and Japan Society for the Promotion of Science.
T. Kishimoto thanks for Special Coordination Funds for Promoting Science
and Technology (Highly Talented Young Researcher) from Japan Science and
Technology Agency.
\end{acknowledgments}

\end{document}